\documentstyle[]{article}
\begin{document}
\title{Quantum correlation functions and the classical limit}
\author{  Charis Anastopoulos \thanks{charis@physics.umd.edu} \\
\small Department of Physics, University of Maryland, College Park, MD20742, USA}
\maketitle

\begin{abstract} 
We study the transition from the full quantum mechanical description of physical systems 
 to an approximate classical stochastic one. Our main tool
 is the identification of the closed-time-path (CTP) generating functional of Schwinger and 
Keldysh with the decoherence functional of the consistent histories approach. 
Given a degree of coarse-graining in which interferences are negligible, we can explicitly write a generating functional for the effective stochastic process 
in terms of the CTP generating functional.
This construction  gives particularly simple results for Gaussian processes. 
The formalism is applied to  simple quantum systems, quantum Brownian motion,  
quantum fields in curved spacetime. Perturbation theory is also explained. 
We conclude with  a discussion on the problem of backreaction of quantum fields in spacetime
 geometry. 
\end{abstract}

\renewcommand {\thesection}{\arabic{section}}   
\renewcommand {\theequation}{\thesection. \arabic{equation}} 
\let \ssection = \section \renewcommand{\section}{\setcounter{equation}{0} \ssection}

\section{Classical vs. quantum probability}
\subsection{Introduction}
The emergence of classical behaviour in quantum systems is a very 
 important question on the foundations of quantum theory. An explanation of how  the 
classical world emerges is absolutely essential for any scheme that has ambitions to go beyond the operational  description of the Kopenhagen interpretation. In recent years the  programme of decoherence has  provided some  insight on how this transition is effected and suggested branches of physics, where relevant phenomena are important, like quantum optics and mesoscopic physics.

From another perspective,  the issue of classicalisation is of significance  in cosmology. 
We want to know  how the perceived classical world is obtained from an underlying description, that is of a (presumably) quantum nature:
 in the early universe,  processes are assumed to be governed by quantum field theory, but later a classical hydrodynamics
 description suffices to capture all relevant physics.  The same question is asked for quantum gravity: only now the focus is 
on the emergence of classical spacetime rather than of the matter fields. 
At a more technical level one is interested to know, when the semiclassical gravity approximation 
(coupling  classical metric variables to quantum fields) is valid.

\bigskip

In all such discussions, the first step  is to establish what is meant by classical behaviour. The notion of classicality can be defined in different ways, according to the context. For instance: 
\\ \\
- the absence of interferences in a given basis: in other words an approximate 
diagonalisation of the density matrix \cite{Zur1,JZ,GKJKSZ}.\\
- determinism or approximate determinism or some form of predictability \cite{Omn8894, GeHa93,Har93a}. \\
- the validity of a hydrodynamic or thermodynamic description for a many-body system \cite{GeHa93, Hal9899,BrHa99}. \\
- the existence of exact or approximate superselection rules \cite{Zur2}. \\ 
\\ \\
Whatever the definition of classicality might be, there is a consensus about how it
 appears.  {\em Coarse-graining} is necessary.
Since the underlying theory is assumed to be quantum theory 
(which is by definition non-classical), one can get a different behaviour only 
by examining a truncated version of the theory. The intuitive picture for emergent classicality is that of a random phase approximation:
 the coarser the description of the 
system, the more the interference phase  cancel out when averaged within the coarse-grained observable.
 The general question is then, which types of coarse-graining can regularly lead to classical behaviour.

In this paper, we take the attitude that a system exhibits classical behaviour, if it admits an
 approximate description in terms of classical probability theory. Since we are interested in systems changing  in time, we ask that {\em the evolution of coarse-grained observables is described by probability theory}: in other words that it should be modeled by a {\em stochastic process}.

  Quantum processes have an important difference 
from stochastic processes: their correlation functions are complex-valued rather than real-valued. 
This is equivalent to the fact that quantum mechanical evolution cannot be described by a probability measure. 
In this paper, 
we focus on how classical correlation functions can be constructed from the quantum
 mechanical ones through coarse-graining, thus providing an effective stochastic description for a quantum system. 
A part of the relevant material has appeared previously 
in \cite{Ana00a}. This presentation is simultaneously an elaboration and a 
simplification of the mathematical constructions performed in this reference, with an eye to possible applications.  

We shall then  apply this formalism in various cases. We will show that in
 Gaussian systems, the classical limit is mostly determined by the {\em real part} of the quantum two-point function. We shall verify this 
in a number of examples: simple harmonic oscillators, the Caldeira-Leggett model of quantum Brownian motion, scalar fields in curved  spacetime. We shall then discuss the perturbation expansion, from which we shall infer that a perturbation expansion of the quantum theory does not imply a perturbation expansion for the corresponding stochastic one. We conclude with a discussion of the validity of the semiclassical approximation in quantum gravity. This is a topic,  which our formalism is particularly adequate to address.

The first step is, however,  a brief summary of classical probability theory.

\subsection{Classical probability}
In classical probability one assumes that at a single moment of time the possible elementary
 alternatives lie in a space $\Omega$, the 
{\em sample space}. Observables are functions on $\Omega$, and are usually called {\em random variables}.

 The outcome of  any measurement can be phrased as a statement that the system is found in 
a given subset $C$ of $\Omega$. Hence, the set of certain 
well-behaved (measurable)  subsets of $\Omega$ is identified with the set of all 
coarse-grained alternatives of the system. To each subset $C$,
 there corresponds an observable $\chi_C(x)$, the characteristic function of the set $C$.
It is defined as $\chi_C(x) = 1$ if $x \in C$ and $\chi_C(x) = 0$ otherwise. 
It is customary to denote the characteristic function of $\Omega$ as $1$ 
and of the empty set as $0$.

Note that  if an observable $f$ takes values $f_i$ in subsets $C_i$ of $\Omega$, we have that 
\begin{equation}
f(x) = \sum_i f_i \chi_{C_i}(x)
\end{equation}
A {\em state} is  intuitively thought of as  a preparation of a system. 
Mathematically it is represented by a measure on $\Omega$, i.e a map that 
to each  alternative $C$, it assigns its probability $p(C)$. 
It has to satisfy the following properties 
\\ \\
- for all subsets $C$ of $\Omega$, $0 \leq p(C) \leq 1$ \\
- $p(0) = 0 ; p(1) = 1$. \\
- for all disjoint subsets $C$ and $D$ of $\omega$, $p(C \cup D) = p(C) + p(D)$ \\ 
\\
Due to (1.1) one can define $p(f) = \sum_i f_i p(C_i)$; $p(f)$ is then clearly the mean value of $f$. The usual notation for the mean value is $\bar{f}$, however the expression $p(f)$ is used, when we want to stress, the state with respect to which the mean value is taken.
When $\Omega$ 
is a subset of ${\bf R}^n$, the probability measures are defined in terms of a probability distribution, i.e. a positive function  on $\Omega$, which 
we shall (abusingly) denote as $p(x)$.
\begin{equation}
p(f) = \int dx \hspace{0.2cm} p(x) f(x)
\end{equation}

There  also exists the notion of {\em conditional probability}: Assume that  in an ensemble
  described by a probability distribution $p(x)$, we measure  whether the property corresponding to the set $C$ is satisfied.
 The subensemble  of all systems that have been found to satisfy this property,
 is then described by the probability distribution $p(x) \chi_C(x)/p(C)$.

Assume now that we have prepared a system in a state $p$ and we 
want to perform a series of measurements of an observable $f = \sum_i f_i \chi_{C_i}(x)$ at time $t_1$ and of $g = \sum_j g_j \chi_{D_j}$
at time $t_2 > t_1$. For simplicity, we shall ignore any self-dynamics of the physical systems as it evolves from $t_1$ to $t_2$.
We can consider a number of measurement situations, labeled by $i$ and $j$, corresponding to an arrangement where the 
filter $C_i$ is  placed at time $t_1$ and the filter $D_j$ at time $t_2$. From a series of measurements one will establish the number of systems in 
the ensemble that pass both filters and hence identify the probability
$p(i,t_1;j,t_2)$, that $C_i$ has been found true at time $t_1$ {\em and then} $D_j$ at time $t_2$ 
\begin{equation}
p(i,t_1;j,t_2) = \int dx p(x) \chi_{D_j}(x) \chi_{C_i}(x)
\end{equation}
Performing this experiment for all different choices of $i$ 
and $j$, we can construct the {\em statistical correlation function}
 
\begin{equation}
\langle f_{t_1} g_{t_2} \rangle = \sum_{ij} f_i g_j p(i,t_1;j,t_2) = p(B A)
\end{equation}
By $BA$ we mean the product of the observables $B$ and $A$, hence $p(BA)$ stands for 
$\int dx \hspace{0.2cm} p(x) B(x) A(x)$.

In general, the system may have intrinsic dynamics. This is implemented by 
  a map $\tau_{t_1,t_2}$ that 
takes the state 
$p(x)$ at time  $t_1$ to the state $ \tau_{t_1, t_2} [p](x)$ at time  $t_2$, 
in such a way as to preserve normalisation and positivity. 
The correlation function then reads
\begin{equation}
\langle f_{t_1} g_{t_2} \rangle = \int dx g(x) \tau_{t_1,t_2}[f p](x)
\end{equation}
Here  $fp$ stands for the state obtained from the multiplication of the function $f(x)$ with the probability distribution $p(x)$.

When we want to study properties of the system at more than one moment of time,
 we need to introduce a sample space for {\em histories}. 
If we denote by $T$ the set of all possible time instants,
 we can identify the space of histories  $\Omega^T$ as a suitable subset of 
$ \times_{t \in T} \Omega_t$, where $\Omega_t$ is a copy of the system's sample space labeled by a moment of time $t$.  The elements of $\Omega^T$ 
are {\em paths} $t \rightarrow x_t$ and will be denoted as $x(\cdot)$.

 A history observable is a function on $\Omega^T$. Given a function $f$ 
on $\Omega$, we can define a family of history observables $F_t$ as 
\begin{equation}
F_t[x(\cdot)]  = f(x(t))
\end{equation}

The state is represented by a probability measure $P$ on $\Omega^T$. It contains information about both initial condition and the dynamics and 
for any function $F$ on $\Omega^T$ it gives its mean value $P(F)$ or simply $\bar{F}$. We can, abusingly, write it in terms of a 
probability distribution on $\Omega^T$  
\begin{equation}
P(F) = \int Dx(\cdot) P[x(\cdot)] F[x(\cdot)]
\end{equation}
The correlation functions $\langle f_{t_1} g_{t_2} \rangle $ can then be written as $P(F_{t_1}G_{t_2})$ in terms of the functions $F_t$ and 
$G_t$ defined by (1.6). The information of the correlation functions of a single observable $f$  is contained in the generating functional 
\begin{eqnarray}
Z_f[J(\cdot)] = \sum_{n=0}^{\infty} \frac{i^n}{n!} \int dt_1 \ldots dt_n \langle f_{t_1} \ldots f_{t_n} \rangle J(t_1) \ldots J(t_n)
 = 
\nonumber \\
 \int Dx(\cdot) P[x(\cdot)]
\exp \left( i \int dt F_t [x(\cdot)] J(t) \right),
\end{eqnarray}
in terms of a function of time $J(t)$, commonly referred to as the "source".

The generating functional is essentially the Fourier transform of the probability measures. 
The definition can be extended for families of observables $f_i$.
Since correlation functions can be operationally determined, it 
is possible, in principle, to determine the probability measure with arbitrary accuracy. 

\subsection{Quantum correlations} 
In the previous section we gave a summary of classical probability theory, thus establishing our notation, and  
 identified the operational meaning of correlation functions in classical probability. 
 
The corresponding structures for a single moment of time
 are well-known in standard quantum theory. Elementary alternatives are rays 
on a complex  Hilbert space $H$, observables are self-adjoint operators 
on $H$, a general property  corresponds to a projection operator and a state to a density matrix. 

Let us now consider an ensemble of quantum systems prepared in a state described by a 
density matrix $\hat{\rho}$ and try to operationally construct the correlation function of two observables $\hat{A} = \sum a_i \hat{P}_i$ and $\hat{B} = \sum_j b_j \hat{Q}_j$
 at times $t_1$ and $t_2 > t_1$ respectively.
Here $\hat{P}_i$ refers to  an exhaustive ($ \sum_i \hat{P}_i = 1$ ) and exclusive ($\hat{P}_i \hat{P}_j = \hat{P}_i \delta_{ij}$) set of projectors,
 and so does $\hat{Q}_j$.

Let the Hamiltonian of the system be $\hat{H}$ and $\hat{\rho}_0$ the state of the system at time $t = 0$. Then a series of measurements will enable us to 
identify the probability that  $\hat{P}_i$ is found true {\em and then } $\hat{Q}_j$ 
is found  true. According to the rules of quantum theory this will be 
\begin{eqnarray}
p(i, t_1; j, t_2) = Tr \left( \hat{Q}_j e^{-i\hat{H}(t_2 - t_1)}\hat{P}_i
 e^{-i\hat{H}t_1 } \hat{\rho}_0 e^{i\hat{H}t_1 } \hat{P}_i e^{i 
\hat{H}(t_2 - t_1)} \right) = 
\nonumber \\ 
Tr \left( \hat{Q}_j(t_2)\hat{P}_i(t_1)\hat{\rho}_0 \hat{P}_i(t_1) \right),
\end{eqnarray}
where we used the Heisenberg picture notation for operators $\hat{A}(t) = e^{i\hat{H}t}\hat{A} e^{-i\hat{H}t}$. 
If we now  vary over all possible values of $i$ and $j$, we can construct the {\em statistical} correlation function between $\hat{A}$ and $\hat{B}$ 
\begin{equation}
\langle \hat{A}_{t_1} \hat{B}_{t_2} \rangle_S = \sum_{ij} a_i b_j  p(i. t_1; j, t_2) 
\end{equation}

But this correlation function is {\em not} what one usually calls correlation function 
in quantum theory. This name is usually employed for the expectation of a product of 
operators
\begin{equation}
\langle \hat{A}_{t_1}\hat{ B}_{t_2} \rangle_Q = Tr \left( \rho \hat{A}(t_1) \hat{B}(t_2) \right) = \sum_{ij} a_i b_j Tr \left( \rho 
\hat{P}_i(t_1)  \hat{Q}_j(t_2) \right).
\end{equation}
This  is a complex-valued object, in contrast to (1.10), that    was constructed using
 probabilities of events and can only 
be real-valued. Then, what does the quantum mechanical correlation correspond to? 
Clearly it is unlike classical correlations. The fact that it is 
complex-valued suggests that it has something to do with quantum mechanical 
quantities such as interference phases. This remark turns out to be accurate. 
In \cite{Ana00b} a scheme was described, in terms
 of which the quantum mechanical correlation functions can be operationally measured. 
It proceeds essentially 
by measuring interference phases between different states. It is a measurement procedure similar to ones used for the Aharonov-Bohm effect  or the 
Berry phase \cite{Ber84}. This is natural in a sense, since the Berry phase is the irreducible element for which quantum theory necessitates the use of complex 
numbers \cite{AnSav00}. However, in the present paper we are interested in the classical limit rather than the full structure of quantum theory and we shall not pursue this topic. The interested reader is referred to \cite{Ana00b}.

\bigskip

We now want to check the possibility that the quantum and the statistical correlation functions coincide. An easily discernible 
case is when $[\hat{A}(t_1), \hat{B}(t_2)] = 0$, i.e. when the measured observabls commute.  More generally, it can  
be verified that a necessary and sufficient  condition is
\begin{equation}
 Re Tr \left( \hat{Q}_j(t_2)\hat{P}_i(t_1)\rho \hat{P}_{i'}(t_1) \right) =  0 ,
\end{equation}
for all $i,j $ and $i' \neq i$. In this case the following property is satisfied 
\begin{equation}
\sum_i p(i,t_i;j,t_2) = Tr(\rho \hat{Q}_j(t_2)) = p(j, t_2)
\end{equation}
for all $j$. This implies that the probabilities assigned to the set of all possible histories satisfy the additivity condition. They, therefore, 
define a classical probability measure. It is evident that in this case the quantum and the statistical correlation functions coincide.

This condition for classicality is exactly the one upon which the formalism of consistent histories is based. This formalism is an indispensable 
part of our analysis and we therefore proceed to examine it next. 

\section{Quantum  processes}
\subsection{Consistent histories}
 
   The consistent histories approach to quantum theory was  developed by 
Griffiths \cite{Gri84}, Omn\'es \cite{Omn8894}, Gell-Mann and Hartle
\cite{GeHa90, GeHa93, Har93a}. The basic object is a  
    {\em history}, which   corresponds to  properties of the physical 
system at successive instants of time. A discrete-time history $\alpha$ will then  
correspond to a string $\hat{P}_{t_1}, \hat{P}_{t_2}, \ldots \hat{P}_{t_n}$ of projectors, 
each labeled by  an instant  of time. From them, one can construct the class operator
\begin{equation} 
\hat{C}_{\alpha} = \hat{U}^{\dagger}(t_1) \hat{P}_{t_1} \hat{U}(t_1) \ldots \hat{U}^{\dagger}
(t_n) \hat{P}_{t_n} \hat{U}(t_n)
\end{equation}
where $\hat{U}(s) = e^{-i\hat{H}s}$ is the time-evolution operator. The probability for the realisation 
of this history is 
\begin{equation}
p(\alpha) = Tr \left( \hat{C}_{\alpha}^{\dagger}\hat{\rho}_0  \hat{C}_{\alpha} \right), 
\end{equation}
 where $\hat{\rho}_0$ is the density matrix describing the system at time $t = 0 $.

But this expression does not define a probability measure in the space of all histories, because  
the Kolmogorov additivity condition cannot
 be satisfied: if $\alpha$ and $\beta$ are exclusive histories, and $\alpha \vee \beta$ 
denotes their conjunction as propositions, then it is not true that 
\begin{equation}
p(\alpha \vee \beta ) = p(\alpha) + p(\beta) .
\end{equation}
 The 
histories formulation of quantum mechanics does not, therefore,  enjoy the status of a genuine
 probability theory.

However,  an
 additive probability measure {\it is} definable, when we restrict to 
particular  sets of histories.
 These are called {\it consistent sets}. They are more conveniently 
defined through the introduction of a new object: the decoherence functional. This is a complex-valued 
function of a pair of histories given by
\begin{equation}
d(\alpha, \beta) = Tr \left( \hat{C}_{\beta}^{\dagger} \hat{\rho}_0 \hat{C}_{\alpha} \right).
\end{equation}
A set of exclusive and exhaustive alternatives is called consistent, if for all  pairs 
of different histories $\alpha$ and $\beta$, we have 
\begin{equation}
 Re \hspace{0.2cm} d(\alpha, \beta) = 0 .
\end{equation}
In that case one can use equation (2.2) to assign a probability measure to this set. The 
consistent histories interpretation then proceeds by postulating that any prediction or 
retrodiction, we can make using  probabilities,   
{\it has always to
 make  reference to a given consistent set}. This  leads to  counter-intuitive   situations, of getting mutually 
incompatible predictions, when reasoning within different consistent sets . 
The predictions of this theory are therefore contextual: but in
 any case, this  is a general feature of  all realist interpretations of quantum theory.

Except for trivial cases, it is only coarse-grained observables that
 satisfy an exact (or approximate) consistency condition. This means 
that the histories are constructed out of projectors $\hat{P}$, whose trace is much larger than unity.

\subsection{The Closed-Time-Path generating functional}
 We saw that in quantum theories probabilities and statistical correlations
 are contained in the decoherence functional; in fact, in its diagonal 
elements. We shall now show that the same is true for the quantum correlation functions. 

Recall that in the decoherence functional projectors enter in a {\em time-ordered 
series}. This suggests that it would be best to use time-ordered 
correlation functions. Let $\hat{A}^a$ denote a family of commuting operators.
 Then the time-ordered two-point correlation function is defined as 
\begin{eqnarray}
G^{2,0}(a_1,t_1;a_2,t_2) = \theta(t_2-t_1) Tr[ \hat{\rho}_0
 \hat{A}^{a_1}(t_1) \hat{A}^{a_2}(t_2)] + \nonumber \\
 \theta(t_1-t_2) Tr[\hat{\rho}_0 \hat{A}^{a_2}(t_2) \hat{A}^{a_1}(t_1) ].
\end{eqnarray}
Here, we have denoted by $\theta(t)$ the step function.

One can similarly define time-ordered $n$-point functions, or anti-time-ordered 
\begin{eqnarray}
G^{0,2}(a_1,t_1;a_2,t_2) = \theta(t_1-t_2)
 Tr[\hat{\rho}_0 \hat{A}^{a_1}(t_1) \hat{A}^{a_2}(t_2) ] + \nonumber \\
 \theta(t_2-t_1) Tr[ \hat{\rho}_0 \hat{A}^{a_2}(t_2) \hat{A}^{a_1}(t_1) ]
\end{eqnarray}
In general, one can define {\em mixed} correlation functions $G^{r,s}$, with $r$ time-ordered and $s$ anti-time-ordered entries, as for instance 
\begin{eqnarray}
G^{2,1} (a_1,t_1; a_2,t_2|b_1, t'_1) = \theta(t_2-t_1) 
Tr[\hat{A}^{b_1}(t_1')\hat{\rho}_0 \hat{A}^{a_1}(t_1) \hat{A}^{a_2}(t_2)] +
\nonumber \\
 \theta(t_1-t_2) Tr[\hat{A}^{b_1}(t_1')\hat{\rho}_0 \hat{A}^{a_2}(t_2) \hat{A}^{a_1}(t_1) ]
\end{eqnarray}
These correlation functions are generated by the Closed-Time-Path (CTP) generating functional associated to the family $A^a$
\begin{eqnarray}
Z_A[J_+,J_-] = \sum_{n,m=0}^{\infty} \frac{{i}^n (-i)^m}{n! m!} \int dt_1 \ldots dt_n dt'_1 \ldots dt'_m  \nonumber \\
G^{n,m}(a_1,t_1; \ldots a_n,t_n | b_1,t'_1;\ldots; b_m,t'_m) 
\nonumber \\
J_+^{a_1}(t_1) \ldots J_+^{a_n}(t_n) J_-^{b_1}(t'_1) \ldots J_-^{b_m}(t'_m)
\end{eqnarray}
here $J^a_+$ and $J^a_-$ are functions of time, that play the role of sources, similar to the ones in equation (1.8)
for the classical stochastic processes.

The name closed-time arose, because in the original conception (by Schwinger
\cite{Schw61}
 and Keldysh \cite{Kel64} the time path one follows is from some initial time 
$t=0$ to $t \rightarrow \infty $, thus covering all time-ordered points and then back from infinity to $0$ covering the anti-time-ordered points.
The total time-path is in effect closed.

Conversely the correlation functions can be read from $Z_A$
\begin{eqnarray}
G_A^{n,m} (a_1,t_1; \ldots ; a_n , t_n | b_1, t'_1; \ldots;b_m, t'_m) \nonumber \\
= (-i)^n i^m \frac{\delta^n}{\delta J^{a_1}_+(t_1) \ldots \delta J^{a_n}_+(t_n)} 
\frac{\delta^m}{\delta J^{b_1}_-(t_1) \ldots \delta J^{b_m}_-(t_m)} Z[J_+,J_-]|_{J_+=J_-=0}. 
\end{eqnarray}

\subsection{Relation between the functionals}

Clearly there must be  a relation between the decoherence functional and the CTP one. One can see in the correlation functions, if we 
assume a single operator  $\hat{A} = \sum_i a_i \hat{P}_i$ and consider a pair of histories
 $\alpha(i_1,t_1; \ldots; i_n,t_n) = \{\hat{P}_{i_1},t_1; \ldots; \hat{P}_{i_n},t_n \} $ 
and $\beta(i_1,t'_1; \ldots;i_n,t'_m) = \{\hat{P}_{j_1},t'_1; \ldots; \hat{P}_{j_m},t_m \}$. Then one can easily verify that 
\begin{eqnarray}
G_A^{n,m}(t_1,\ldots, t_n;t'_1,\ldots, t'_m) = \sum_{i_1\ldots i_n} \sum_{j_1 \ldots j_m} a_{i_1}\ldots a_{i_n}
 b_{j_1} \ldots b_{j_m} \nonumber \\
\times d[\alpha (i_1,t_1; \ldots; i_n,t_n),\beta (j_1,t_1; \ldots; j_m,t_m)]
\end{eqnarray}

The straightforward relation is nonetheless not possible to show in an 
elementary fashion. One  needs to consider correlation functions at all times $t$ and this 
necessitates a description in terms of histories that can have temporal support over the whole of the real line,
or at least a continuous subset of it.
This can be achieved in the framework 
of  continuous-time histories \cite{IL95,ILSS98, Sav99a,SavAn00}.
However, this   requires a significant upgrading of the 
formalism of quantum mechanical histories. The key idea 
 is to represent histories by projectors on a tensor product of Hilbert spaces
 $\otimes_{t \in T} H_t$ \cite{I94} in analogy to the 
construction of the history sample space classically. A suitable Hilbert
 space (not  a genuine tensor product) can be constructed \cite{IL95}
for the case that $T$ is a continuous set and the decoherence functional 
can be defined as a bilinear, hermitian functional on this space. 
It can then be shown that as a functional it is essentially a double
 "Fourier transform" of the CTP generating functional. 

This proof is to be found in \cite{Ana00a} and is elementary once one 
follows the logic of the construction. Here we shall restrict ourselves 
to a convenient statement of this result.

 Let us assume that we have a family of commuting self-adjoint operators
 $\hat{A}^i$.
 Their spectrum is then a subset 
$\Omega$ of some vector space ${\bf R}^n$.  
Any operator that commutes with  $\hat{A}^i$ is in one-to-one correspondence  to functions $f(x)$
  with  $x \in \Omega$ and can be written as $f(\hat{A})$. Like the classical 
case we can construct a space of histories $\Omega^T$ as a suitable subset of $\times_{t \in T} \Omega_t$. Subsets of  
of $\Omega^T$ are  histories of the quantum mechanical observables $\hat{A}^i$. 

 The decoherence functional is then a map that to each pair of 
subsets $C$ and $C'$ 
of $\Omega^T$ it assigns a complex number in such  a fashion that the following properties are satisfied \cite{Har93a,IL94}
\\ \\
- $d(C',C) = d^*(C,C')$, {\em hermiticity} \\
- $d(0,C) = 0$, {\em null triviality}\\
- $d(1,1) = 1$, {\em normalisation}\\
- $d(C \cup C', D) = d(C,D) + d(C',D)$ for disjoint $C$ and $C'$, {\em additivity}.
\\ \\
 Such a decoherence functional 
 can be constructed as a continuum-limit of the discrete-time expressions (2.2). Because of the additivity 
condition, one  can  {\em formally} write the decoherence functional as 
an integral over $\Omega^T \times \Omega^T$
\begin{equation}
d(C,D) = \int Dx(\cdot) Dx'(\cdot) \hspace{0.15cm} \Delta[x(\cdot)|x'(\cdot)]  \hspace{0.15cm} \chi_C[x(\cdot)] \chi_D[x'(\cdot)],
\end{equation}
in terms of a function $D: \Omega^T \times \Omega^T$, that plays the role of an integration kernel. This is in complete 
analogy to the stochastic probability measure $P[x(\cdot)]$ of equation (1.7). 

One can view $\Delta[x(\cdot)| x'(\cdot) ]$ as 
the decoherence functional between 
a pair of fine-grained histories $x(\cdot)$ and $x'(\cdot)$, only that such histories cannot be represented by projectors
 on a Hilbert space. For 
example, if these histories where defined on the configuration space for  the time interval $[t_i,t_f]$, one could 
write the standard expression \cite{Har93a}
\begin{equation}
\Delta[x(\cdot)|x'(\cdot)] = \rho_0[x(t_i),x'(t_f)] \delta[x(t_f),x'(t_f)] \hspace{0.12cm} e^{i S[x(\cdot)] - i S[x'(\cdot0]}, 
\end{equation}
in terms of the matrix elements of the initial density matrix $\rho_0$, the standard configuration space action and a delta-funtion for 
the final-time points of the paths. 

\bigskip

If $Z_A[J_+,J_-]$ is the CTP generating functional associated to $\hat{A}^i $ we have 
\begin{equation}
Z_A[J_+,J_-] = \int Dx(\cdot) Dx'(\cdot) e^{i \int dt J^a_+(t) x^a(t)} e^{- i \int dt J_-^a(t) x'^a(t)} \Delta[x(\cdot)|x'(\cdot)]
 \end{equation}
In other words viewed as a bi-functional over the functions on $\Omega^T$ the 
decoherence functional is identical to the CTP generating functional. 
The only difference is on the  type of functions upon which they take values - 
the first on characteristic functions, the second on complex valued functions of 
unit norm. In fact, equation (2.12) 
amounts to
\begin{eqnarray}
G^{n,m}(a_1,t_1; \ldots; a_n,t_n| b_1, t_1; \ldots; b_m,t_m) = \int \int Dx(\cdot) Dx'(\cdot) \nonumber \\
 \times x^{a_1}(t_1) \ldots x^{a_n}(t_n)
x'^{b_1}(t'_1) x'^{b_m}(t_m) \Delta[x(\cdot)|x'(\cdot)] 
\end{eqnarray} 

Hence there exists the following correspondence between classical and quantum probability \\
\\  \\ 
\begin{tabular}{llcr}
{}             & {\em Quantum} &  $ \rightarrow $ & {\em Classical} \\
{\em Probabilities }&    $d(C,C')$& $\rightarrow$& $p(C)$ \\
{\em Correlations }&    $Z[J_+,J_-]$&$\rightarrow$ & $Z^{cl}[J]$
\end{tabular}
\\ \\  \\
The probability measure is a real-valued functional on functions
 of  $\Omega^T$, while the decoherence functional is a {\em hermitian} 
bilinear functional on the functions of $\Omega^T$. In a
 given system, one goes from $d$ to $p$, when the decoherence condition (2.5) 
is satisfied, while in both cases one goes from probabilities to correlations
 through a Fourier transform.

What we will next show is how to effect the transition from the CTP generating
 functional to a stochastic process for  the coarse-grained variables.
 Working at the level of the correlation functions makes the construction of 
stochastic differential equations easier than working at the level of probabilities.

\section{ From quantum to classical}
\subsection{The basic choice for coarse graining}
In order to study the transition from quantum to classical, we need to 
choose the variables, upon which we shall concentrate. This amounts to a 
 choice of a family $\hat{A}^i$ of intercommuting operators. Now, a {\em maximal} family
 of intercommuting operators generically contains full information about the evolution of 
the quantum 
system. (Possible exceptions to this rule are trivial cases,  as, for instance, when 
the Hamiltonian and the initial density matrix  commutes with all $\hat{A}^i$). 

One can implement a coarse-graining procedure even at this stage. It suffices to 
take for $\hat{A}^i$ a non-maximal family of operators. This is the case, 
for instance, in quantum Brownian motion models. If we assume that the total system 
consists of a large number of harmonic oscillators, a maximal family of intercommuting 
operators consists of the position operators of all particles. When we choose to focus 
on a single one of them, we effectively coarse-grain by treating the remaining  degrees of freedom as an environment.
 This is the type of coarse-graining associated with the studies of environment-induced decoherence.

However, this type of coarse-graining does not suffice. 
One has usually to consider smeared values of the relevant observables. This
 is effected by considering projectors, which are   sufficiently smeared over $\Omega$. 
We shall take  $\Omega$ to be   ${\bf R}^s$ so 
its points will be vectors $x^a$.

 In general, it is difficult to work with characteristic  
functions, so we will work with {\em smeared characteristic functions}. 
If we denote by $|x|$ their Euclidean distance,
 then a good choice for the projector is the function
\begin{equation} 
f_{\bar{x}}(x) = \exp ( - \frac{1}{ 2 \sigma^2} |x - \bar{x}|^2 ) 
\end{equation}
This Gaussian is not a sharp projector; it is strongly peaked in 
 a sphere of length $\sigma$ around the point $\bar{x}$,
 hence it is a good approximation to a true projector for not very large values of $\sigma$. 

We can construct  discrete-time histories, consisting  of 
 projectors $f_{\bar{x}_{t_i}}$ centered around $\bar{x}_{t_i}$, at each time $t_i$ . One such  history  can be viewed as 
a discretised approximation to a coarse-grained history in continuous-time, centered around
 a path $t \rightarrow \bar{x}(t)$. 

We now  consider two such discretised histories,
 centered at the same time points $t_i$ each corresponding to a different path 
$\bar{x}(\cdot)$ and $\bar{x}'(\cdot)$. Let us denote them by 
$\alpha_{\bar{x}(\cdot)}$ and $\alpha_{\bar{x}'(\cdot)}$.
 If we expect our system to exhibit classical behaviour, then  
the off-diagonal elements of the decoherence functional will fall rapidly, 
whenever $\delta^2 = || \bar{x}(\cdot) - \bar{x}'(\cdot)||^2:=
\sum_i |\bar{x}_{t_i} - \bar{x}'(t_i)|^2$ is much larger than $ N \times \sigma^2$. (Here $N$ 
is the number of time-steps).
Typically one has 
\begin{equation}
d(\alpha_{\bar{x}(\cdot)}, \alpha_{\bar{x}'(\cdot)}) = O(e^{- \delta^2/ N \sigma^2}),
\end{equation}
or some other type of rapid fall-off. For pure initial states, this behaviour is expected, {\em when 
$\sigma^2$ much larger than the uncertainties of the initial state and the Hamiltonian evolution 
preserves this property \cite{DoHa92}}.
In this case, the diagonal elements are close to defining 
probabilities for coarse-grained histories centered around $\bar{x}(\cdot)$ and with a spread $\sigma$ 
at each moment of time. 

Now, we want to find a probability distribution that  would give these  values for the probabilities
 of these histories. A single-time projector is centered in a volume of $\Omega$ of size 
\begin{equation}
\int dx f_{\bar{x}}(x) = ( 2 \pi \sigma^2)^{r/2}  .
\end{equation}
A probability distribution on the space of 
(discretised) paths that 
reproduces  these expressions for probabilities of these coarse grained sets 
is  
\begin{equation}
p[\bar{x}(\cdot)] = \frac{1}{ (2 \pi \sigma^2)^{rn/2}} d( \alpha_{\bar{x}(\cdot)}, \alpha_{\bar{x}(\cdot)}) ,
\end{equation}
where $n$ is the number of time-steps assumed. 
(Dividing by the volume turns the probabilities
of events into a density.) 

One can use equation (2.12) to write 
\begin{eqnarray}
d(\alpha_{\bar{x}(\cdot)}, \alpha_{\bar{x}(\cdot)}) = 
\int Dx(\cdot) Dx'(\cdot) \nonumber \\ 
\times 
\exp( - \frac{1}{2\sigma^2} ||x(\cdot) - \bar{x}(\cdot)||^2 - \frac{1}{2\sigma^2} 
||x'(\cdot) - \bar{x}(\cdot)||^2) \Delta[x(\cdot)| x'(\cdot)]
\end{eqnarray}
Note that  our expressions are still defined with respect to discrete time. 

From equation (2.13) we see that the kernel $\Delta$ can be obtained from the inverse Fourier 
transform of the CTP generating functional. This yields 
\begin{eqnarray}
p[\bar{x}(\cdot)] =  \int DJ_+(\cdot) DJ_-(\cdot) \! e^{-1/4 \pi
 (J_+\cdot J_+ + J_- \cdot J_-) - i/\sqrt{2 \pi} \bar{x} \cdot 
(J_+ - J_-) } 
\nonumber \\ 
\times Z[J_+/(\sqrt{2 \pi}\sigma), J_-/(\sqrt{2 \pi} \sigma)]   
\end{eqnarray}

There is now no multiplicative term, that  depends on the number of time steps. Hence, 
 one can safely go to the continuum limit from this expression. 
We have denoted as $J \cdot x = 
\sum_i J(t_i) x(t_i)$ and at the continuum limit this expression will become an 
integral.

To get the generating functional for the classical correlation functions one Fourier transforms 
the probability measure to get (we now drop the index that refers to our choice of variables for the correlation functions)
\begin{equation}
\tilde{Z}^{cl}[J] = e^{- \frac{\sigma^2}{8} J \cdot J} \int dR(\cdot) e^{- 1/{2 \pi} R \cdot R}
 Z[\frac{R}{\sqrt{2 \pi} \sigma}+\frac{J}{2},\frac{R}{\sqrt{2 \pi} \sigma} - \frac{J}{2}] ,
\end{equation}
with $R = \frac{1}{2}(J_+ + J_-)$. 

This generating functional  needs to be normalised to unity by assuming that
$\tilde{Z}^{cl}[0] = 1$. The normalisation condition is not kept, because we have employed approximate 
characteristic functions. Had we used a sharp characteristic function, 
the construction would automatically guarantee normalisation. Now there is  a deviation from unity of the  order of $\sigma^2$.

The expression (3.7) can be simplified.
 Assume that we have a classical stochastic process for the variables 
$x(\cdot)$, with a generating functional $Z_0[J]$. Let us follow the same procedure for 
coarse-graining as before, using the approximate projectors (3.1). The coarse-grained 
generating functional would be 
\begin{equation}
\tilde{Z}^{cl}[J] = e^{-\frac{\sigma^2}{2} J \cdot J} Z_0[J]
\end{equation}

This means, that we can consider the generating functional  in equation (3.7) as coming from coarse-graining  
a classical stochastic process, with twice the degree of coarse graining as the one from quantum 
theory. One can then drop the term outside the integral in (3.7) as coming from coarse-graining 
of an underlying  stochastic process given by 

\begin{equation}
Z^{cl}[J] = \int dR(\cdot)  e^{- 1/{2 \pi} R \cdot R}
 Z[\frac{R}{\sqrt{2 \pi} \sigma}+ \frac{J}{2},\frac{R}{\sqrt{2 \pi} \sigma} - \frac{J}{2}]
\end{equation}

This equation gives the stochastic correlation functions in the classical limit of the quantum 
system described by the CTP generating functional $Z[J_+,Z_-]$. It should be always kept in mind,
that this process 
 {\em gives reliable results only on scales much larger than $\sigma^2$}. 

We should now pause for a minute and examine the 
assumptions we used in order to arrive here. 

First, one should ask what is the meaning of the parameter $\sigma$. Is it arbitrary or not? 
In principle it is not. It is the  degree of coarse-graining, that is necessary  in order that the fall-off (2.5) 
of the off-diagonal elements of the decoherence functional is manifested. It, therefore, has to be much 
larger than the natural scale associated with microscopic processes; however,  it ought to be small compared to macroscopic scales.
 
In principle, $\sigma$ can be determined from a full study 
of the decoherence functional. Usually a measure of coarse-graining is the trace of the corresponding 
positive operator. However,  we have here considered only commuting variables with continuous spectrum and 
the trace of the operators corresponding to (3.7) is infinite.  This is related to the fact that there is no default 
{\em universal} scale, by which to judge whether $\sigma$ is large. This problem is remedied 
by considering phase space coarse-grainings as we shall see shortly. In this case the natural 
scale is $\hbar = 1$ and one can say that $\sigma^2 >> 1$ in order to have consistency of histories.
It should nonetheless be much smaller than a macroscopic time scale 
by which we observe phenomena.

More precisely ,  the stochastic approximation (3.9) is accurate within an order  of $(l_{mic}/ \sigma)^2$, where 
$l_{mic}$ is the microscopic scale that is determined by the dynamics or the initial state. However, there 
is also an error proportional to $(\sigma/L_{mac})^2$, where $L_{mac}$ is the macroscopic scale of 
{\em observation}, i.e. the scale of accuracy we are interested in having. This is due to the use of the 
Gaussian approximation for the projectors. 
Overall we have an error of the order of 
\begin{equation}
c_1 (l_{mic}/ \sigma)^2 + c_2 (\sigma/ L_{mac})^2,
\end{equation}
where $c_1$ and $c_2$ are constants of the order of unity. It is, therefore, evident that a separation 
of scales is {\em necessary}, if the stochastic description is to make any sense. 

Second, the general logic of this construction is to identify a stochastic process that adequately 
describes the evolution of the classicalised coarse-grained observables. There is a subtle difference 
from the consistent histories scheme, in that we do not seek to construct consistent sets for 
the system and hence make statements about individual quantum systems. Our approach is more operational.
Given that quantum theory is a model that provides the statistical behaviour of physical systems, we ask 
to construct a different model based on probability theory that describes some regime of the same physical 
system. For this purpose we utilise the consistency condition in order to identify the validity of our 
approximation. 
Then, we build the probability distribution from the diagonal elements of the decoherence functional. 

\subsection{Phase space coarse grainings}
One does not have to restrict to correlation functions of a family of commuting operators 
in order to construct the CTP generating functional. By considering correlation functions in 
both position and momentum, it is possible to generalise the definition (2.9). Indeed in this equation 
it is not necessary to assume that the operators $\hat{A}^a$ are commuting. We assumed 
commutativity, because we wanted a description in terms of paths on the common spectrum of these operators.

However, in any Hilbert space, that carries a representation of the canonical commutation 
relations
\begin{equation}
[\hat{q}^i, \hat{p}^j] = i \delta^{ij}
\end{equation}
it is possible to assign a  function on the phase space $\Gamma = \{(q^i,p^i)\}$ for each operator 
by means of the Wigner 
transform 
\begin{eqnarray}
\hat{A} \rightarrow F_A(q,p) = \int d \xi d \chi e^{-iq \xi - i p \chi} 
Tr \left( \hat{A} e^{i\hat{q} \xi + i \hat{p} \chi} \right) : = Tr \left( \hat{A} \hat{\Delta}(q,p)\right)
\end{eqnarray}
An important property of this transform is that it preserves the trace: 
\begin{equation}
Tr \hat{A} = \int dq dp F_A(q,p)
\end{equation}
However, the Wigner transform does not preserve multiplication of operators. The defining condition $\hat{P}^2 = \hat{P}$ 
for projectors is therefore not preserved and a projector is mapped into some general positive 
function, rather than a characteristic function of a subset of $\Gamma$.

Since any operator can be represented by a function of $\Omega$, histories would be represented 
by functions on a space $\Gamma^T$, which will be a suitable subspace of $\times_{t \in T} \Gamma_t$.
In reference \cite{Ana00a} it was shown that a decoherence functional can be constructed as an
hermitian bilinear functional on the space of functions on $\Gamma^T$. 
And it is related to the CTP generating functional by means of a Fourier transform. 

All the formulas in the previous paragraph can then be reinterpreted to fit the phase space 
context, by allowing the variables $x$ to denote both $q$ and $p$. (The dimension of $\Gamma$ is clearly 
even).
 The main difference is 
that the Gaussian function $f_{\bar{x}}$ corresponds to an operator $\hat{F}$ with a {\em finite} 
trace. By virtue of (2.1) and (2.13)
\begin{equation}
Tr \hat{F} = (2 \pi \sigma^2)^{r/2}  
\end{equation}
The parameter  $\sigma$ has units of {\em action} and is an {\em absolute} measure of the degree of coarse-graining on phase space.
Consistency occurs whenever $\sigma^2 >> \hbar$, where $\hbar$ provides the natural length 
scale on phase space. In fact, in the study, of a large class of closed quantum systems, Omn\'es has 
showed \cite{Omn89} that the off-diagonal elements of the decoherence functional are of the  order of 
$(\hbar/\sigma)^{r/4}$, where $r = 2k$ is the dimension of $\Gamma$.  Hence, even if $\hbar/ \sigma \sim 
10^{-8}$ there is a substantial degree of decoherence to justify the use of classical probability 
 and {\em $\sigma$ is still sufficiently 
small} compared to some external macroscopic scales to justify the use of the Gaussian approximation for the projector. From a macroscopic perspective it would
 be sufficient to consider the leading order in $\sigma^2$ of the correlation functions.

The study of phase space histories is more intricate, because one has to choose proper 
units for position and momentum, by which to write Euclidean norm in the coarse-grained projector (2.1).
For classicality it is not only necessary to have a large value of $\sigma$, but 
 the {\em choice of units has to be 
preserved by the dynamical evolution} \cite{Omn89,Ana99}. This is a non-trivial condition that largely depends 
on the system's Hamiltonian. For this we shall prefer to employ configuration space 
coarse-grainings.

Whenever we have a representation of the canonical commutation relations we can 
define the coherent states 
\begin{equation}
|z \rangle = |\chi \xi \rangle = e^{i\hat{q} \xi + i \hat{p} \chi}  |0 \rangle,
\end{equation}
where $| 0 \rangle$ is a fiducial vector, often taken to be the lowest energy 
eigenstate. The important point is that one can assign to a  large class of 
density matrices $\hat{\rho}$  a function $f_{\rho}(\chi,\xi)$ (its P-symbol)
 defined by
\begin{equation}
\hat{\rho} = \int d \chi d \xi f(\chi, \xi)| \chi \xi \rangle \langle \chi \xi|.
\end{equation}

If one then denotes by $Z_{\chi_0 \xi_0}[J_+, J_-]$ the CTP generating functional
 corresponding to an initial state given by $|\chi_0 \xi_0 \rangle$ then the 
CTP generating functional for the same system but a different initial state $\hat{\rho}$
\begin{equation}
Z[J_+,Z_-] = \int d \chi_0 d \xi_0 f_{\rho}(\chi_0, \xi_0) Z_{\chi_0 \xi_0}[J_+, J_-]
\end{equation}
and a similar equation would hold   for the classical limit,  provided that 
the degree of coarse-graining necessary for decoherence is determined by the study
 of the state $\hat{\rho}$ rather than the coherent states.

\subsection{ Gaussian processes}

Let us now consider a quantum system described by a Gaussian CTP generating functional. Its most general form would be 
\begin{eqnarray}
Z[J_+,J_-] = \exp \left( - \frac{i}{2} J_+ \cdot L \cdot J_+ + \frac{i}{2}
J_- \cdot \bar{L} \cdot J_- \right.
\nonumber \\
\left. + i J_+ \cdot K \cdot J_- + i (J_+ - J_- ) \cdot X \right)
\end{eqnarray}

Here we have denoted by $L$ kernels of the 
form $L^{ab}(t,t')$, by $J \cdot L \cdot J' = \int dt dt' J^a(t) L^{ab}(t,t') J^b(t')$
 and the bar denotes complex conjugation. $X$ denotes the one-point function $G^{10} = G^{01}$ and 
\begin{eqnarray}
i L^{ab}(t,t') = G^{2,0}(a, t; b, t') - X(a,t) X(b,t') \\
i K^{ab}(t,t') = G^{1,1}(a,t|b,t') + G^{1,1}(b,t'|a,t) - 2 X(a,t) X(b,t')
\end{eqnarray}

We can write $ L = L_1 - i L_2$ and $K = K_1 - i K_2$, in terms of 
the real-valued kernels $L_1,L_2, K_1, K_2$. The hermiticity condition on the CTP generating functional would then entail 
\begin{eqnarray}
L_1^T = L_1 \hspace{3cm} L_2^T &=& L_2 \\
K_1 = 0 \hspace{3cm}   K_2 &=& 2 L_2
\end{eqnarray}

Evaluating the integral (3.7) yields 
\begin{equation}
Z^{cl}[J] = e^{ - J \cdot \Xi \cdot J + i J \cdot X}
\end{equation}
where 
\begin{equation}
\Xi =     L_2  + \frac{1}{4\sigma^2} L_1 \cdot L_1 
\end{equation}

It is worth noticing, that whenever the term $L_2$ is dominant, the classical 
two-point function is independent of the coarse-graining scale and equal to the real part 
of the quantum two-point function. However, this simplification can occur only in  Gaussian systems.
\section{Examples}
\subsection{Harmonic oscillators}
For a single harmonic oscillator with frequency $\omega$ and  mass $m$ in a thermal state,
 we have for the configuration space correlation functions 
\begin{eqnarray}
L_1(t,t') &=& -\frac{1}{2 m \omega} \sin \omega|t-t'| \\
L_2(t,t') &=&  \frac{1}{2 m \omega} \coth (\beta \omega/2) \cos\omega (t-t')  \\
\end{eqnarray}
and equation (3.24) gives $\Xi(t,t')$.

One should recall that the smearing scale $\sigma$ is determined 
by the condition (2.5) on the fall-off of the diagonal elements of the 
decoherence functional. Here $\sigma^2$ should be much larger than $(2 m \omega)^{-1}$, the 
position uncertainty of the {\em ground}
 state. This can be verified by direct evaluation, but it is made plausible by the following 
observation: a thermal state has a positive P-symbol, and hence its quantum behaviour 
is identical to the one of the coherent states, which in a Gaussian system is identical with 
that of the vacuum. 

The term $L_1 \cdot L_1$ is proportional to $ (\sigma^2 m \omega)^{-1}$, hence 
comparatively small. In particular, 
 at high temperature $\beta \omega << 1$ the $L_2$ term is  dominant, the 
correlation function is $\sigma$-independent and one recovers the classical result.

Let us recall that this system does not describe a harmonic oscillator in contact with a heat bath; it describes a {\em closed} 
system, evolving unitarily and  prepared in a thermal state (whatever that might mean).
Physically more relevant  is the case of an oscillator undergoing quantum 
Brownian motion, to be taken up later. 

But we shall first examine the case, where the system is initially 
prepared in a squeezed  state.  A squeezed state $| r, \phi \rangle$ 
is the zero  eigenstate of the operator
\begin{equation}
\hat{b} = \cosh r/2 \hat{a} + \sinh r/2 e^{i \phi} \hat{a}^{\dagger}, 
\end{equation}
where $r \geq 0$.
The correlation function $L_1$  is identical to the one for the vacuum case,  while 
\begin{eqnarray}
L_2(t,t') = \frac{1}{2m \omega} \left( \cosh r \cos \omega (t-t') +
 \sinh r \cos \omega (t + t' - \phi) \right)
\end{eqnarray}

Clearly it is necessary that $\sigma^2 >> \frac{\cosh r}{2 m \omega}$ in order to have 
decoherence. In that case the $L_1^2$ term is again negligible. For values of $r$ at the order 
of unity, it is not different from the vacuum case, but for large $r$ the degree of coarse-graining 
necessary for classicality might become too large to allow us to obtain any useful information.
This is what is meant, when we say that squeezed states are highly non-classical states. 

\subsection{Quantum Brownian motion}
We shall study here the Caldeira-Leggett model \cite{CaLe83, HPZ92, HaYu96}, i.e. a single harmonic oscillator of mass $M$ 
and frequency $\omega$ coupled linearly to a bath of harmonic oscillators in a thermal state.
More precisely  the system is defined  by the Hamiltonian
\begin{eqnarray}
\hat{H} = \frac{\hat{p}^2}{2M} + \frac{1}{2} M \omega^2 \hat{x}^2 + 
\hat{x} \sum_i c_i \hat{q}_i 
+ \sum_i ( \frac{\hat{p}_i^2}{2m_i} + \frac{1}{2}m_i \omega_i^2 \hat{q}_i^2 )  .
\end{eqnarray}
From the Heisenberg equations of motion we get 
\begin{eqnarray}
M \frac{d^2}{dt^2}  \hat{x} + M \omega^2 \hat{x}^2 &=& - \sum_i c_i \hat{q}_i , \\
\frac{d^2}{dt^2}  \hat{q}_i + \omega_i^2 \hat{q}_i^2 &=& - \frac{c_i}{m_i} \hat{x} .
\end{eqnarray}
The second equation has a solution 
\begin{equation}
\hat{q}_i(t) = \hat{q}_{0i} \cos \omega_i t + \frac{\hat{p}_{0i}}{m_i \omega_i} \sin \omega_i t 
- \frac{c_i}{m_i \omega_i} \int_0^t ds \sin \omega_i(s-s') \hat{x}(s),
\end{equation}
which when substituted into (4.7) yields
\begin{equation}
\frac{d^2}{dt^2} \hat{x} + \omega^2 \hat{x} - \frac{2}{M} \int_0^t ds \eta(t-s) \hat{x}(s)
 = -\frac{1}{M} \sum_i c_i \hat{q}_i(t) 
\end{equation}
Here 
\begin{equation}
\eta(s) = \sum_i \frac{c_i^2}{2 m_i \omega_i} \sin \omega_i s ,
\end{equation}
is known as the {\em dissipation kernel}. Let us denote by $u(t)$ the solution of the homogeneous equation corresponding to (4.10) with 
the initial conditions $u(0) = 1$ and $\dot{u}(0) = 0$. It can be identified as the inverse 
Laplace transform of the function 
\begin{equation}
\tilde{u}(s) = \frac{1}{s^2 + \omega^2 - 2/M \tilde{\eta}(s)},
\end{equation}
where $\tilde{\eta}$ is the Laplace transform of the dissipation kernel. 
We can then write the solution of (4.10) as 
\begin{eqnarray}
\hat{x}(t) = \hat{x}_0 u(t) + \frac{\hat{p}_0}{M} \dot{u}(t) 
\nonumber \\
- \frac{1}{M} \sum_i c_i [ \hat{q}_{0i} \int_0^t ds u(t-s) \cos \omega_i s + 
\frac{\hat{p}_{0i}}{m_i \omega_i} \int_0^t ds u(t-s) \sin \omega_i s ]
\end{eqnarray}
Now we assume that the initial state of the system is factorisable to a thermal state
 at temperature 
$T = \beta^{-1}$ for the environment and a density matrix $\hat{\rho_0}$ for the distinguished 
oscillator. In this case, we can easily see that the expectation value 
\begin{equation}
x(t) = Tr[\hat{\rho}_0 \hat{x}(t)] = x_0 u(t) + \frac{p_0}{M} \dot{u}(t),
\end{equation}
is a solution of the dissipative equations of motion , while the two-point function reads 
\begin{eqnarray}
Tr [\hat{\rho}_0 \hat{x}(t) \hat{x}(t') ] = (\Delta x_0)^2 u(t) u(t') + \frac{(\Delta p_0)^2}{M^2} \dot{u}(t) 
\dot{u}(t') + 
\nonumber \\
\frac{C_{pq}}{M} [u(t) \dot{u}(t') + u(t') \dot{u}(t)] + \frac{i}{2M} 
[u(t) \dot{u}(t') - u(t') \dot{u}(t)] \nonumber \\
+ \frac{1}{M^2} [\int_0^t ds \int_0^{t'} ds' u(s) \nu (s-s') u(s')  +  i 
\int_0^t ds \int_0^{t'} ds' u(s) \eta (s-s') u(s')  ]
\end{eqnarray}
Here $\Delta x_0, \Delta p_0$ and $C_{pq}$ are the uncertainties and correlation between 
position and momenta at $t=0$. Also, 
\begin{equation}
\nu(s) = \sum_i \frac{c_i^2}{2 m_i \omega_i} \coth \frac{\beta \omega_i}{2} 
\cos \omega_i s 
\end{equation}
is known as the {\em noise kernel}.

From this equation, it is easy to determine the kernels $L$ and $K$. If we write 
the last line in equation (4.15) as $\frac{1}{M^2} [N(t,t') + i H(t,t')]$ we have 
\begin{eqnarray}
L_1(t,t') = \theta(t-t') [\frac{1}{M}[u(t) \dot{u}(t') - u(t') \dot{u}(t)] +
\frac{1}{M^2} H(t,t')]
\nonumber \\
+ \theta(t'-t) [\frac{1}{M}[u(t') \dot{u}(t) - u(t) \dot{u}(t')] + \frac{1}{M^2} H(t',t)] \\
L_2(t,t') = (\Delta x_0)^2 u(t) u(t') + \frac{(\Delta p_0)^2}{M^2} \dot{u}(t) 
\dot{u}(t') \nonumber \\
+ \frac{C_{pq}}{M} [u(t) \dot{u}(t') + u(t') \dot{u}(t)] + \frac{1}{M^2} N(t,t')
  \end{eqnarray}

Now we want  to derive the stochastic limit, to which these quantum correlation functions correspond.
We shall use equation (3.24). Note however, that the necessary degree of coarse-graining in order to 
achieve decoherence must be in a phase space region much larger than the one occupied by the 
initial state. Hence we take the coarse-graining scale to be much larger than the uncertainties and 
correlations of the initial state. This allows us to drop all terms but $N(t,t')$ in equation (4.15).
The semiclassical equations are then largely independent of the details of the initial state.

To see this in more detail we need to specify a given distribution of modes and couplings in the 
environment. This information is encoded in the spectral density 
\begin{equation}
I(k) = \sum_{i} \frac{c_i^2}{2 m_i \omega_i} \delta(k - \omega_i)
\end{equation}
If the spectral density is specified, then we can fully determine the noise and dissipation kernels as 
\begin{eqnarray}
\eta(s) = \int dk I(k) \sin ks \\
\nu(s) = \int dk I(k) \coth(\frac{\beta k}{2}) \cos ks . 
\end{eqnarray}

A large class of physically interesting choices for spectral density are 
\begin{eqnarray}
I(k) = \frac{2 M \gamma}{\pi} k \left(\frac{k}{\Theta} \right)^{s-1} \hspace{1cm}, k &\leq& \Lambda  \nonumber \\
 0 \hspace{1cm}, k &>& \Lambda,
\end{eqnarray}
where $\Lambda$ is a high-frequency cut-off. The exponent $s$ determines the infra-red behaviour 
of the bath. For $s=1$ the environment is called ohmic, for $s < 1$ subohmic and for $s > 1$
supraohmic. 

\paragraph{High temperature} 
For high temperature $\beta \Lambda << 1$, the $L_2$ term is proportional to $\beta^{-1}$ and 
dominates the classical correlation function. The classical stochastic process will be then {\em independent} 
of the precise choice of coarse-graining. The two point function for $x$ will then be
\begin{equation}
\Xi(t,t') = \frac{1}{ M^2} \int_0^t ds u(t-s)\int_0^{t'}ds' u(t'-s') \nu(s-s')
\end{equation}
 
But this is the correlation function for the solution of the 
classical stochastic differential 
equation
\begin{equation}
M\frac{d^2}{dt^2} x(t) + M\omega^2 x(t) - \int_0^t ds \eta(t-s) x(s) = f(t),
\end{equation}
where $f(t)$ is a Gaussian process with two-point function
\begin{equation} 
\langle f(t) f(t') \rangle = \eta(t-t').
\end{equation} 
The noise kernel then gives the correlation function for an external noise perturbing the 
classical dissipative equations of motion. This justifies its name and recover results 
suggested by path integral techniques \cite{HPZ92}, or explicitly  proved only in particular regimes \cite{Har93a,DoHa92}.
 {\em Note, however, 
that this is true only when the $L_2$ term in (3.24) dominates, as is the case of high 
temperature}. In the previous paragraph it is by necessity that the $L`^2_1$ term is small. Here 
it is not the case, because $L_1$ contains also a contribution from the environment degrees 
of freedom that might give a substantial contribution in certain regimes. 
In the general case the $L_1^2$ term might be of importance, something 
that implies that  the 
stochastic limit will not be given by such a simple expression 
in terms of an external force guided by the noise kernel. 

 \subsection{Scalar field}
Consider now the case of a free, massive scalar field in  vacuum. The correlation functions 
read 
\begin{eqnarray}
L_1({\bf x},t;{\bf x'},t') = - \int \frac{d^3k}{(2 \pi)^3}\frac{1}{2 \omega_{{\bf k}}} 
e^{-i{\bf k \cdot x}} \sin \omega_{{\bf k}} |t-t'| \\
 L_2 ({\bf x},t;{\bf x'},t') =  - \int \frac{d^3k}{(2 \pi)^3} \frac{1}{2 \omega_{{\bf k}}} 
e^{-i{\bf k \cdot x}} \cos \omega_{{\bf k}} (t-t')
\end{eqnarray}
Since $\omega_{{\bf k}} \geq m$,  coarse-grainings with $\sigma^2 m^2 >> 1$  for each mode will manifest a 
suppression of the interferences. As in the harmonic oscillator the $L_1^2$ term, will 
be negligible and the $L_2$ will provide the classical correlation function. 

For a massless field, there is no natural coarse-graining scale for all modes. One needs to 
take larger values of $\sigma$ to adequately deal with the 
 infra-red modes. Configuration space coarse-graining 
is clearly bad in such a  case. The analysis  has to be performed in phase space and 
 provides the same results as the $m \neq 0$ case \cite{Blen91}. 

Consider now the case of a free scalar field in a general globally hyperbolic spacetime.
 The reader is referred to the standard treatments of Birrell and Davies \cite{BiDa82} and Wald \cite{Wald94}. 
Let us by $t$ denote the time coordinate in a spacelike foliation, by ${\bf x}$ the spatial 
coordinates and let us use a collective index $\alpha$ to label the modes. The  Heisenberg-picture field will read then 
\begin{equation}
\hat{\phi}({\bf x},t) = \sum_{\alpha} \left[ \hat{a}_{\alpha} u_{\alpha}({\bf x},t) + \hat{a}^{\dagger}_{\alpha}
\bar{u}_{\alpha}({\bf x}, t) \right] ,
\end{equation}
where $u_{\alpha}({\bf x},t)$ are some complex-valued solutions to the Klein-Gordon equation and 
\begin{equation}
[\hat{a}_{\alpha}, \hat{a}^{\dagger}_{\beta}] = \delta_{\alpha \beta}.
\end{equation}

Let us assume that the system is found in a Gaussian state $|\Omega \rangle$, which is annihilated by the 
operator $\hat{a}_{\alpha}$. The action  of $\hat{a}^{\dagger}$ on $| \Omega \rangle$ produces 
all states of the Fock space. It is easy to compute 
\begin{equation}
\langle \Omega|\hat{\phi}({\bf x},t) \hat{\phi}({\bf x'},t')| \Omega \rangle =  
\sum_{\alpha} \bar{u}_{\alpha}({\bf x'},t') u_{\alpha}({\bf x},t)
\end{equation} 

Let us now consider a
particular instant $ t = 0 $ as reference time, in which  the mode functions 
$u^0_{\alpha}({\bf x}) = u_{\alpha}({\bf x},t)$ form an orthonormal basis 
\begin{eqnarray}
\int d {\bf x} u^0_{\alpha}({\bf x}) u^0_{\beta}({\bf x}) = \delta_{\alpha \beta} \\
\int d {\bf x} \bar{u}^0_{\alpha}({\bf x}) u^0_{\beta}({\bf x}) = 0.
 \end{eqnarray}
Here, we wrote as $d {\bf x}$ the  volume element of the push-backed spacetime metric  
in the spacelike surface $t = 0 $.
Any scalar function on a instant of time can be decomposed in modes 
\begin{equation}
u_{\alpha}({\bf x},t) = A_{\alpha \beta}(t) 
u^0_{\beta}({\bf x}) + B_{\alpha \beta}(t) \bar{u}^0_{\beta}({\bf x}) 
\end{equation}
The matrices $A$ and $B$ are the Bogolubov coefficients and satisfy the matrix identity 
\begin{equation}
A^{\dagger} A - B^{\dagger} B = 1,
\end{equation}
which essentially means that time evolution is given (classically) by a symplectic transformation.
 The correlation function (4.30) 
reads then 
\begin{eqnarray}
\langle \Omega|\hat{\phi}({\bf x},t) \hat{\phi}({\bf x'},t')| \Omega \rangle =  \nonumber \\
\bar{u}^0({\bf x'}) A^{\dagger}(t') A(t) u^0({\bf x}) + u^0({\bf x'}) B^{\dagger}(t') B(t)
 \bar{u}^0({\bf x})  \nonumber \\
 + \bar{u}^0({\bf x'}) A^{\dagger}(t') B(t) \bar{u}^0({\bf x}) + u^0({\bf x'}) B^{\dagger}(t') A(t) u^0({\bf x}),
\end{eqnarray} 
where a matrix notation has been employed.

This gives for the kernels
\begin{eqnarray}
L_1(t,{\bf x};t',{\bf x'}) &=& - 2 \theta(t-t') \hspace{0.1cm} Im \hspace{0.1cm}\bar{u}^0({\bf x'})[A^{\dagger}(t') A(t) - B^{\dagger}(t') B(t) ] u^0({\bf x}) \nonumber \\
&-& 2 \theta(t'-t) \hspace{0.1cm}
 Im \hspace{0.1cm} \bar{u}^0({\bf x})[A^{\dagger}(t) A(t') - B^{\dagger}(t) B(t') ] u^0({\bf x'})  \\
L_2((t,{\bf x};t',{\bf x'}) &=& 2 \hspace{0.1cm} Re \hspace{0.1cm} 
\bar{u}^0({\bf x'})[A^{\dagger}(t') A(t) 
+  B^{\dagger}(t') B(t) ] u^0({\bf x}) \nonumber \\
&+& 2 \hspace{0.1cm} Re \hspace{0.1cm}  \bar{u}^0({\bf x'}) A^{\dagger}(t') B(t) \bar{u}^0({\bf x})
\end{eqnarray}

The Bogolubov transformation is a generalisation of the squeezing transformation we studied earlier. For  a single mode we had to coarse-grain in regions of configuration space much larger than the uncertainty. It is similar in this case. We can coarse-grain each mode separately at a scale $\sigma^2$. From equation (4.35) we can read the uncertainty for each mode (defined using  the mode functions $u^0_{\alpha}$). This will be equal to 
\begin{equation}
(\Delta \phi_{\alpha})^2 =  [A^{\dagger}A + B^{\dagger}B]_{\alpha \alpha}(t)
+[A^{\dagger}B + B^{\dagger}A]_{\alpha \alpha}(t)
\end{equation}

The first term is always larger in norm than the second due to Schwarz inequality. 
The degree of  coarse-graining in  each mode must be much larger than  $(A^{\dagger}A + B^{\dagger}B)_{\alpha \alpha}(t)$ for all $t$.
 To have a uniform coarse-graining scale for all modes it is necessary that $\sigma^2$ 
 has to be much larger than the norm of the matrix
\begin{eqnarray}
|| A^{\dagger} A + B^{\dagger} B ||(t) = ||2 A^{\dagger} A - 1 || (t) < 2 ||A||^2(t)
\end{eqnarray}
The matrix $A$ has a finite norm be virtue of the Bogolubov identity (4.34).  Now, in order to
 have a meaningful coarse-graining procedure it is necessary that the norms 
of the Bogolubov matrices must be bounded in time. Hence one can
 write a sufficient condition for the possibility of a coarse-graining scale $\sigma^2$
 valid for all modes is that 
\begin{equation}
\sigma^2 >> \sup_t ||A(t)||^2
\end{equation}
We can then employ equation (3.24) to identify the correlation function  for the
 classicalised field. Due to (4.40) the $L_1^2$ term will in general be smaller, as in the case of the 
squeezed system hence 
the distribution (4.37) gives the correlation function of the classical stochastic process.

\subsection{Perturbative expansion}

So far we have considered only the case of Gaussian processes. We shall now study the case of non-quadratic systems through the use of perturbation theory. Let us by $Z_0[J_+,J_-]$ denote a CTP generating functional that can be exactly evaluated, e.g. a Gaussian. In general, a perturbative expansion around $Z_0$ will be of the form 
\begin{equation}
Z[J_+,J_-] = \exp \left(i  F[i\frac{\delta}{\delta J_+}, - i \frac{\delta}{\delta J_-}] \right) Z[J_+,J_-],
\end{equation}
  in terms of a functional $F[x(\cdot),x'(\cdot)]$, that depends on some coupling constant.  
In the case that the CTP generating functional depends on configuration space variables,
 the Hamiltonian is of the form $\hat{H}_0 + V(\hat{x})$ and the initial state is the vacuum, then we have 
\begin{equation}
 Z[J_+,J_-] = \exp \left( i \int dt V[i\frac{\delta}{\delta J_+(t)}] 
- i V[-i\frac{\delta}{\delta J_-(t)}] \right) Z_0[J_+,J_-]
\end{equation}
Substituting $J_{\pm} = R/(\sqrt{2 \pi} \sigma) \pm J/2$ in equations (4.42) 
 yields for the classical generating functional
\begin{eqnarray}
Z^{cl}[J] = \int dR(\cdot) e^{-\frac{1}{2 \pi} R \cdot R} \nonumber \\
\exp \left( i \int dt 
(V[ \frac{i}{2} \frac{\delta}{\delta J(t)} + i \sqrt{\pi/2} \sigma \frac{\delta}{\delta R(t)}] \right. \nonumber \\
\left. - V[ \frac{i}{2} \frac{\delta}{\delta J(t)} + i \sqrt{\pi/2} 
\sigma \frac{\delta}{\delta R(t)}])\right) Z_0[\frac{R}{\sqrt{2 \pi} \sigma} +
\frac{J}{2},\frac{R}{\sqrt{2 \pi} \sigma} -\frac{J}{2}]
\end{eqnarray}
If we assume that the length scale by which the potential varies is much larger than $\sigma$
 we can keep the lower 
order term in $\sigma$ in the exponential to get 
\begin{eqnarray}
Z^{cl}[J] = \int dR(\cdot) e^{-\frac{1}{2 \pi} R \cdot R} \hspace{3cm}
 \nonumber \\
\times \exp \left( - \int dt \sqrt{\pi/2} 
\sigma V'[\frac{i}{2} \frac{\delta}{\delta J(t)}] \frac{\delta}{\delta R(t)} \right)
Z_0[R,J] \nonumber \\
\simeq  Z_0^{cl}[J] - \int dt \frac{1}{2 \sqrt{2 \pi}} 
\sigma V'[\frac{i}{2} \frac{\delta}{\delta J(t)}] \int dR(t) R(t) e^{-\frac{1}{4 \pi} R \cdot R} 
Z_0[R,J] ,
 \end{eqnarray}
where $Z_0[J]$ is defined from $Z_0[J_+,J_-]$ via (3.9) and we wrote for brevity 
$Z_0[R,J] = Z_0[R/\sqrt{2 \pi} \sigma+J/2, R/\sqrt{2 \pi} \sigma - J/2]$. 

 The above expression is the leading 
order in $\sigma$ of the generating functional and is valid only {\em when the potential V 
is assumed to vary in macroscopic scales}. 

Note, that as we see from (4.44), a perturbation expansion of the quantum theory does 
{\em not} generically 
amount to a perturbative expansion in the corresponding classical limit. These results are of relevance for the study 
of perturbation theory in quantum Brownian motion \cite{HuPaZh93, Brun93}.

\section{Concluding remarks}

Let us now summarise our results. We first showed that 
 the quantum mechanical correlation functions 
do not correspond to the statistical properties of a physical system and hence do not correspond 
to a classical stochastic process. Then we  explained the relation of Schwinger and Keldysh's 
 CTP generating functional to the decoherence functional of the consistent histories approach 
to quantum theory. This enabled us to use the decoherence condition for histories in order 
to develop a procedure of going from a quantum process to a stochastic process that corresponds 
to a given degree of coarse-graining for a class of observables. The end result was 
equation (3.9) that gives the relation between the classical generating functional and the CTP 
generating functional. But we should keep in mind that any results of the stochastic
 description that give some detailed structure in scales smaller than $\sigma$ 
are  unreliable. 

We then proceeded to study examples. We showed that in Gaussian processes the classical limit 
is a Gaussian process, with a correlation function given by the {\em real part } of the quantum 
two-point function plus a term that depends on the coarse-graining scale. The second term is 
negligible in general, but this is not necessary in the case of quantum Brownian motion at low 
temperatures, because then it largely depends on properties of the bath, rather than the 
initial condition of the distinguished system. 

\bigskip

Now, what is the use of these results? The answer is that a stochastic description might be more 
amenable to our intuition than a quantum one. For instance, given a stochastic process, one can 
simulate the evolution of individual systems and identify typical behaviours. This is something 
that {\em cannot be done} in quantum theory. We have no (uncontroversial) way to describe the random 
evolution of an individual system's observables, while in classical probability we can 
write equations of the Langevin type. In this sense, the stochastic approximation 
captures an aspect of the quantum mechanical randomness that might be manifested on macroscopic or 
even intermediate scales. Hence for Brownian particle, the knowledge of the noise due to the environment 
might be  sufficient for certain applications. 

Substituting the full quantum behaviour for an approximate description in terms of Langevin 
equation is necessary in cases, where the quantum system acts as a driving force term for another 
 classical one. This is the case, for instance, of any detector coupled to a quantum 
field. While one often uses a quantum description for a detector, this is clearly inappropriate 
for realistic systems: a detector is a macroscopic system and the description of its effective 
behaviour in terms of a simple Schr\"odinger equation is an idealisation that is 
hardly  justified from first principles.
On the other hand a detector can be treated as a classical system - e.g. with respect to its center 
of mass- that is coupled to a stochastic driving force, that arises from the fluctuation of the 
quantum system, with which it is coupled. 
 In this case the stochastic description of a quantum system is not only 
convenient, but necessary.

The situation is similar in the case of quantum field theory in curved spacetime as far as the issue 
of backreaction is concerned. The quantum field acts as a source for the classical gravitational field 
via its stress-energy tensor. One writes then the semiclassical gravity equation 
\begin{equation}
G_{\mu \nu} = \kappa \langle T_{\mu \nu} \rangle
\end{equation}
There is an underlying assumption in any use of this equation. As it stands it is nonsensical: 
on the left hand side is an observable for an individual system and on the right hand side 
an ensemble average. What is implied is that  i) the behaviour of the quantum field is in 
some regime approximately deterministic and ii) the corresponding classical value of the 
stress energy tensor is equal to the expectation value over the field's state. 

Of course these two assumptions need to be verified. In order for the quantum field to behave 
classically, coarse-graining is necessary. Even with coarse-graining it is not necessary  that 
the system will classicalise. For instance, in the one-dimensional case,  time evolution might cause
a continuous increase of the squeeze parameter  in a given mode. In this case no fixed 
degree of coarse-graining is good at all times and it is very little we can do with a classical 
description. Let us assume, however, that this is not the case. According to our earlier analysis, this 
would mean that we can choose a coarse-graining scale satisfying equation (4.40). Then, one might 
expect that the system will exhibit stochastic behaviour for at least some of its properties. 

In order for assumption i) to be valid, the stochastic process for the stress-energy tensor 
should have small deviations from its mean. This is rarely true, as in simple spacetimes 
one can show that the quantum  fluctuations of the stress-energy tensor are of the same order 
of magnitude as its mean value \cite{HuPh00b}. This leads to the point 
 that 
the semiclassical description of backreaction of quantum fields onto geometry ought to have 
a stochastic component \cite{Hu89,CaHu94,HuSi95,HuMa95,
VeCa96, MaVe99a, MaVe99b}. 

This is where our results in section 4.3  are of relevance. First, we argued that the condition 
(4.40) 
is sufficient to obtain classicality; then we showed that it is essentially the real part of 
the two-point function of the fields that gives the two-point function at the classical limit. Now the stress energy tensor is a quadratic functional of the fields 
\begin{equation}
T^{\mu \nu} (x) = \frac{1}{2} \nabla^{\mu}\phi(x) \nabla^{\nu} \phi(x) - \frac{1}{2} 
g^{\mu \nu} \phi(x)(-\nabla_{\rho} \nabla^{\rho} + m^2) \phi(x)
\end{equation}

Its expectation value can be read from the two-point function
via point splitting and renormalisation. This means that we can define 
\begin{eqnarray}
\langle T^{\mu \nu}(x) \rangle  = \lim_{x' \rightarrow x}
  \frac{1}{2} [ 1/2 (\nabla^{\mu}_x \nabla^{\nu}_{x'} + \nabla^{\mu}_x \nabla^{\nu}_{x'}) \nonumber \\
-  1/2[g^{\mu \nu}(x) + g^{\mu \nu}( x')](\nabla^{\rho}_x \nabla_{ x' \rho} + m^2) \Delta_2(x,x')
\end{eqnarray}

Now, the imaginary point of the two point function vanishes as $t \rightarrow t'$ as can 
be easily checked by equation (4.35). Since this part is antisymmetric the 
symmetrisation of the derivatives  in the definition   (5.3) implies 
that it vanishes. Hence,  the 
 {\em quantum mechanical expectation of the stress energy tensor equals the 
classical stochastic expectation }. However, as we said before, one needs to take the higher 
order correlations into account in order to have a consistent backreaction. 
In this case, the naive prescription of using the quantum mechanical correlation
 functions of the stress energy tensor breaks down. The correct higher 
order correlation functions for the stress energy tensor ought to be constructed 
from the classical correlations (4.37). And clearly the stochastic process for the stress energy tensor is {\em not} Gaussian, even though it is obtained from the Gaussian process for the field $\phi$.

The general conclusion in this description is that the backreaction of 
quantum fields to geometry can be described by a  stochastic differential equation 
of the type 
\begin{equation}
G_{\mu \nu} = \kappa T_{\mu \nu}[\phi]
\end{equation}
 where $T^{\mu \nu} $ is a random variable, a functional of the classicalised field 
$\phi(\cdot)$ that is defined by a Gaussian stochastic process with generating functional 
given by (4.37). This result is conditional upon (4.40) holding that defines the possibility
 of having a robust coarse-graining.
\\
\\
\\
{\bf Acknowledgments:} Research was supported partly by NSF and partly by the Onassis Foundation.

\end{document}